\documentclass[12pt]{article}

\usepackage{amsmath}
\usepackage{amssymb}
\usepackage{mathrsfs}
\usepackage{epsfig}
\usepackage{latexsym}
\usepackage{amssymb}
\usepackage{amsmath}
\usepackage{epsfig}
\usepackage{natbib}
\usepackage[dvips]{color}

\def\Tr{\textrm{Tr}}

\begin{document}

\newcommand{\bea}{\begin{eqnarray}}
\newcommand{\eea}{\end{eqnarray}}

\begin{titlepage}

\setcounter{page}{1} \baselineskip=15.5pt \thispagestyle{empty}

\begin{flushright}
ULB-TH/08-18\\ 
\end{flushright}
\vfil

\begin{center}
{\LARGE {\bf Membranes on  Calibrations}}
\end{center}
\bigskip\

\begin{center}
{Chethan KRISHNAN\footnote{\texttt{Chethan.Krishnan@ulb.ac.be}} 
and Carlo MACCAFERRI\footnote{\texttt{cmaccafe@ulb.ac.be}} }
\end{center}

\begin{center}
\textit{{\it Physique Th\'eorique et Math\'ematique,
International Solvay
Institutes, \\ Universit\'e Libre de Bruxelles, ULB Campus Plaine C.P. 
231, 
B--1050 Bruxelles,
Belgium}}
\end{center} \vfil

\noindent 
M2-branes can blow up into BPS funnels that end on calibrated intersections of M5-branes. In this quick note, we make the observation that the constraints required for the consistency of these solutions are automatic in Bagger-Lambert-Gustavsson (BLG) theory, thanks to the fundamental identity and the supersymmetry of the calibration. We use this to explain how the previous {\em ad hoc} fuzzy funnel constructions emerge in this picture, and make some comments about the role of the 3-algebra trace form in the derivation.

\vspace{0.5in}

\begin{center}
\end{center}

\vfil

\end{titlepage}


\pagestyle{headings}


\section{\bf Introduction}

The difficulty with M-theory 
is that essentially all the things we know about it are 
indirect\footnote{See \cite{willy} for an exception to this.}. We know 
that its low energy limit is 11 dimensional supergravity, we know some of 
its dual string theories under various compactifications, and we know that 
its solitonic extended objects are membranes and five-branes. But unlike 
in the case of string theory D-branes, the lack of a worldsheet 
description prevents us from explicitly constructing the low energy field 
theory living on these branes. 

But indirect arguments can sometimes go a long way. We can construct (say) 
membrane solutions in the 11 D supergravity and try to deduce the 
properties of the M2-worldvolume theory by investigating the SUGRA 
solution. First of all, since the brane breaks only half of the 32 
supersymmetries of 11 D supergravity, we expect that the theory has 16 
supersymmetries. Next, we can take a near horizon limit which corresponds 
to looking at the low energy theory on the worldvolume. The 
result of the near-horizon limit is an $AdS_4 \times S^7$ background. The 
isometries of $AdS_4$ correspond precisely to the conformal group in 2+1 
dimensions, and the isometries of $S^7$ should
give rise to an $SO(8)$ R-symmetry. In short, we expect that the theory on 
the 
worldvolume of M2-branes is a maximally supersymmetric superconformal 
theory in 2+1 dimensions with an $SO(8)$ R-symmetry.

But for a while, nobody knew how to build such a theory and in fact it was 
shown that conventional approaches were doomed to fail \cite{schwarz}. The 
situation changed recently, when Bagger \& Lambert \cite{BL1} 
and Gustavsson \cite{Gus1} managed to construct such a 
theory using ideas inspired by non-associative algebras. This theory is 
the first ingredient in our work. 

The second ingredient consists of BPS funnels constructed using M-branes. 
Fuzzy funnels are static configurations were one set of branes expands 
into another. In the case of string theory, such solutions have been 
written down explicitly for the case where D1-branes expand into a single 
D3-brane (BIon) \cite{bion} and into multiple intersecting 
D3-branes \cite{const-lamb}. This latter case can be interpreted also as 
D1-branes ending on a calibrated cycle wrapped by the D3 \cite{Gibbons}. In all these 
cases the configurations are solutions of the BPS equations (also called 
Nahm equations) for the low energy theory on the D1-branes which is two 
dimensional maximally supersymmetric Yang-Mills with $U(N)$ gauge group.

The generalization of the BIon to the case of the M-theory branes was done 
first by Basu and Harvey \cite{Basu}. They wrote down an equation which 
was an {\em ad hoc} generalization of the BPS equation, without an 
underlying supersymmetric action to start with. But it captured the 
expected features of an M2-brane expanding into an M5. This construction 
was instrumental in the recent progress in M-brane theory. The Basu-Harvey 
equation was generalized to the case of multiple intersecting M5s by 
Berman and Copland in \cite{bermcop}. They found that for general 
calibrations the configuration has to satisfy some additional constraints 
(invisible in the Basu-Harvey case) if the BPS equation had to be 
compatible with the equation of motion. 

When the Bagger-Lambert-Gustavsson theory came along, one of its 
interesting aspects was that it could reproduce the Basu-Harvey equation 
directly as a BPS equation. The question we address in this paper is how 
the more general BPS funnels of \cite{bermcop} fit into the theory. In 
fact, we will see that the analogues of the constraints found in 
\cite{bermcop} for general calibrations are automatic in the BLG theory 
because of the structure of the 3-algebra and supersymmetry. Therefore, the
previous fuzzy funnel solutions can be constructed in the BLG 
framework by relating the bracket structures. We also make some comments 
about the conditions on the trace form of the 3-algebra if we want to 
derive the BPS equations by following a Bogomol'nyi ansatz on the energy 
functional. In particular, we discuss the Lorentzian trace forms discussed recently in the literature \cite{ABC}.

Recent work on M2-branes include investigations of their moduli spaces
\cite{Jacques}, the relation of M2 branes to D2 branes 
\cite{Christopher}, uniqueness theorems of the 3-algebra 
\cite{BPG}, and explanations for the 
uniqueness by looking at the boundary of the open membrane version of the theory 
\cite{boundary}.
Related references are \cite{stuff}. Useful reviews on the pre-BLG state of 
the art on M-branes and their interactions are \cite{Berman}.

\section{\bf M2-brane Field Theory}

The Lagrangian for Bagger-Lambert-Gustavsson theory is written using the 
``structure constants" $f^{abc}{}_d$ of a 3-algebra. For the moment, they 
can be thought of as 
coefficients that dictate the couplings between the internal components 
of various fields. With this understanding the Lagrangian is written as
\bea
\label{BLG}
\nonumber {\cal L} &=& -\frac{1}{2}D^\mu X^{aI} D_\mu X^{I}_{a}
+\frac{i}{2}\bar\Psi^a\Gamma^\mu D_\mu \Psi_a
+\frac{i}{4}\bar\Psi_b\Gamma_{IJ}X^I_cX^J_d\Psi_a f^{abcd}\\
&& - V(X)+\frac{1}{2}\varepsilon^{\mu\nu\lambda}\left(f^{abcd}A_{\mu
ab}\partial_\nu A_{\lambda cd} +\frac{2}{3}f^{cda}{}_gf^{efgb}
A_{\mu ab}A_{\nu cd}A_{\lambda ef}\right).
\eea
Upper case Latin indices denote the transverse directions of the M2-brane 
($I,J, K=3,...,10$),  and lower case denotes internal indices in the 
3-algebra. Greek indices denote the worldvolume directions ($\mu, \nu, 
\lambda=0,1,2$). It is useful to think of the Gamma matrices as those of 
11 dimensions, constructed as tensor products of $SO(2,1)$ and $SO(8)$ 
Gamma matrices. The scalar potential takes the explicit form $V(X) = \frac{1}{12} f^{abcd}f^{efg}{}_d X_a^I X_b^J X_c^K X_e^I X_f^J X_g^K.$
The covariant derivative is $D_\mu X^I_d =\partial_\mu X^I_d - \tilde 
A_\mu{}^{c}{}_{d}X^I_c$, where $\tilde A_{\mu}{}^c{}_d = A_{\mu 
ab}f^{abc}{}_{d}$.
The gauge field ensures that the supersymmetry algebra can be closed, but 
it has no propagating degrees of freedom. This is as it should be because 
supersymmetry forbids more bosons. On shell, the Lagrangian above is 
closed under the following SUSY variations
\bea
\nonumber \delta X^I_a &=& i\bar\epsilon\Gamma^I\Psi_a\\
\delta \Psi_a &=& D_\mu X^I_a\Gamma^\mu \Gamma^I\epsilon
-\frac{1}{6}
X^I_bX^J_cX^K_d f^{bcd}{}_{a}\Gamma^{IJK}\epsilon \label{susy}\\
\nonumber \delta\tilde A_{\mu}{}^b{}_a &=& i\bar\epsilon
\Gamma_\mu\Gamma_IX^I_c\Psi_d f^{cdb}{}_{a},
\eea
if the structure constants satisfy the so-called fundamental identity:
$f^{efg}{}_{d}f^{abc}{}_{g}=f^{efa}{}_{g}f^{bcg}{}_{d}
+f^{efb}{}_{g}f^{cag}{}_{d}+f^{efc}{}_{g}f^{abg}{}_{d}$.
In the above, $\epsilon$ is the 32 component real spinor of 11 dimensions, 
which satisfies the condition $\Gamma_{012}\epsilon=\epsilon$ because the 
other 16 supersymmetries are broken by the M2s. This chirality condition 
picks out only the $\Gamma_{012}\Psi=-\Psi$ components in the superpartner 
fermions. The equations of motion that follow from the closure of the SUSY 
variations (and also from the action) are,
\bea
\label{EOMS}
\nonumber\Gamma^\mu D_\mu\Psi_a
+\frac{1}{2}\Gamma_{IJ}X^I_cX^J_d\Psi_bf^{cdb}{}_{a}&=&0\\
 D^2X^I_a-\frac{i}{2}\bar\Psi_c\Gamma^I_{\ J}X^J_d\Psi_b f^{cdb}{}_a
   -\frac{\partial V}{\partial X^{Ia}}    &=& 0 \\
\nonumber \tilde F_{\mu\nu}{}^b{}_a
  +\varepsilon_{\mu\nu\lambda}(X^J_cD^\lambda X^J_d
+\frac{i}{2}\bar\Psi_c\Gamma^\lambda\Psi_d )f^{cdb}{}_{a}  &=& 0.
\eea 
with the gauge field strength
$
\tilde F_{\mu\nu}{}^b{}_a  =\partial_\nu \tilde A_{\mu}{}^b{}_a  -
\partial_\mu \tilde A_{\nu}{}^b{}_a-\tilde A_{\mu}{}^b{}_c\tilde 
A_{\nu}{}^c{}_a
+ \tilde A_{\nu}{}^b{}_c \tilde A_{\mu}{}^c{}_a$.

When investigating fuzzy funnel solutions we will not be interested in 
gauge fields and fermions. In this case it is often instructive to write 
the action in terms of the 
3-algebra directly instead of using explicit internal components:
\bea
\label{action}
{\cal L}_B=-\frac{1}{2}\Tr\,(\partial_\mu X^I,\partial^\mu X^I)
-\frac{1}{12}{\rm Tr}([X^I,X^J,X^K],[X^I,X^J,X^K]) 
\eea
This is to be understood in terms of generators $T^a$ ($a=1,...,N$) such 
that $X^I \equiv X^{I}_a\,T^a$. These generators satisfy 
$[T^a,T^b,T^c] = f^{abc}{}_{d}\,T^d$,
and have a trace form $h^{ab}=\Tr\,(T^a,T^b)$.
The trace form can be used to raise indices in $f^{abc}{}_{d}$. In fact to 
write down the above action (\ref{BLG}), we have implicitly assumed the 
existence of such a trace form and assumed also that the $f^{abcd}$ 
constructed that way is fully antisymmetric in all indices. In terms of 
the generators, this anti-symmetry condition means that $\Tr(T^a,[T^b,T^c,T^d]) = -\Tr([T^a,T^b,T^c],T^d)$.
It should be emphasized that if we forget about the action and merely want 
to close the supersymmetry variations, we only need the 
fundamental identity and neither the existence of the trace form nor the 
full antisymmetry. For future use, we also write down the fundamental 
identity in terms of the 3-algebra generators:
\bea
\label{FID}
[T^a,T^b,[T^c,T^d,T^e]] &=&[[T^a,T^b,T^c],T^d,T^e]
+[T^c,[T^a,T^b,T^d],T^e]\\
\nonumber && +\ [T^c,T^d,[T^a,T^b,T^e]],
\eea

\noindent
\section{Nahm Equations and Calibrations}


We now proceed to construct the BPS equations for fuzzy funnels in Bagger-Lambert-Gustavsson theory following a Bogomol'nyi ansatz. Similar constructions are well known in the 
literature \cite{bermcop}, the point here is only that we would like to work exclusively 
with the 3-algebra structure. 

We are interested in static solutions, so we start with the Hamiltonian 
for  (\ref{action}):
\bea
\label{H}
E=\frac{1}{2} \int d^2 \sigma \left( {\rm Tr}(\partial_\sigma 
X^I,\partial_\sigma X^I)+
\frac{1}{3!} {\rm Tr}([X^I,X^J,X^K],[X^I,X^J,X^K])  \right)
\eea
where $\sigma_2$ can be gauge-fixed to be $X^2$. The aim is to write this 
as
\bea
E=\frac{1}{2}\int d^2 \sigma \left( {\rm Tr}\left(\partial_\sigma 
X^I-\frac{g_{IJKL}}{3!}
[X^J,X^K,X^L]\right)^2+T \right)
\eea
where $T$ is a total derivative, so that we can identify the perfect 
square piece with a BPS equation\footnote{Note that this assumes implicitly 
that the trace form in the 3-algebra is positive definite. This means that we are working with the 3-algebra ${\cal A}_4$ with structure constants $\epsilon^{abcd}$ in this section \cite{BPG}.}. By direct 
computation, it turns out that in order to do this we need the constraint 
\bea
\label{constraint}
\frac{1}{3!}g_{IJKL}g_{IPQR}{\rm Tr}([X^J,X^K,X^L],[X^P,X^Q,X^R])=  
{\rm Tr}([X^I,X^J,X^K],[X^I,X^J,X^K]) 
\eea
to be satisfied. If this holds, then we can conclude that the BPS 
configurations of the theory are given by 
\bea
\label{Nahm}
\partial_\sigma X^I-\frac{g_{IJKL}}{3!}[X^J,X^K,X^L]=0.
\eea

These equations are the Bagger-Lambert-Gustavsson version of the 
(generalized) Nahm equations. The coefficients $g_{IJKL}$ are determined 
by the specific calibration on which the the M2-branes end. Notice that 
the conventional Nahm equations are written in terms of matrix 
commutators \cite{const-lamb}, but here the brackets stand for the 3-product of the 
3-algebra. 

The coefficients $g_{IJKL}$ characterize the specific BPS solution under 
consideration. In practice this corresponds to having M-brane 
configurations that leave some of the supersymmetry unbroken. For BLG 
theory, from (\ref{susy}), setting $ \delta \Psi=\partial_\mu X^I\Gamma^\mu \Gamma^I\epsilon
-\frac{1}{6} [X^I,X^J,X^K]\Gamma^{IJK}\epsilon $
to zero results in 
\bea
\sum_{I<J<K}[X^I,X^J,X^K]\Gamma^{IJK}(1-g_{IJKL}\Gamma^{IJKL 
2})\epsilon=0,
\eea
This can be solved as in \cite{const-lamb, bermcop} by defining 
projectors. We will not repeat the details, the end result is that once we 
have a set of mutually commuting set of supersymmetries, we can rewrite 
the above equation  in the form
\bea
\label{susycond}
\sum_{g_{IJKL}=0}[X^I,X^J,X^K]\Gamma^{IJK}\epsilon=0.
\eea
The sum is over ${I,J,K}$ such that $g_{IJKL}=0$. This can be re-expressed 
as a set of conditions to be satisfied by the 3-brackets. 

Now we will show that the supersymmetry of the calibration, together with 
the fundamental identity is enough to show that the constraint equations 
are satisfied in Bagger-Lambert-Gustavsson theory. Together with the 
identification of the 3-algebra structure with the Nambu 4-bracket 
proposed in \cite{Basu, bermcop}, this will imply that the fuzzy funnels 
on these calibrations are naturally thought of as solutions of BLG. The 
$g_{IJKL}$ have a direct correspondence to the calibrating forms of the 
cycle on which the M5 wraps, and therefore are fully antisymmetric in their 
indices. When $g_{IJKL}=\epsilon_{IJKL}$, we end up with Basu-Harvey and 
the constraints are identities. For other calibrations $g_{IJKL}$ considered in \cite{bermcop} with more $X^I$ active, things are a bit more complicated. But it turns 
out that for each of them, after using the conditions arising from 
supersymmetry (\ref{susycond}), we can write the constraints 
(\ref{constraint}) in terms of the 3-algebra as
\bea
\label{weird}
[[X^L,X^{[I},X^J],X^{K]},X^L] = 0, 
\eea
\vspace{-0.4in}
\bea
[[X^M,X^{[I},X^J],X^K,X^{L]}]=0.\label{note}
\eea
Here $^{[\ , \ , \ ]}$ stands for antisymmetrization, and the activated $I,J,K,L$ depend crucially on the chosen calibration. But the general structure is all we need to demonstrate our point. The first of these relations follows immediately upon setting $a=e$ (or equivalently, symmetrizing in those indices) in (\ref{FID}). After a bit of massaging, the second relation can also be shown to hold due to the fundamental identity. Some useful relations are collected in the Appendix. 

There is another constraint that we need to check for full consistency of 
the solution. This arises from the BLG equations of motion (\ref{EOMS}). 
With the gauge field and the fermion set to zero, the remaining constraint 
comes from the $A_{\mu}{}^b{}_a$ equation of motion:
\bea
X^I_a\partial_\sigma X^I_b f^{abc}{}_d=0.
\eea
This can be reinterpreted as the statement that the 3-bracket $[X^I,\partial_\sigma X^I,Z]$ vanish for arbitrary $Z$. If we use the generalized Nahm equation (\ref{Nahm}) to rewrite $\partial_\sigma X^I$, this becomes
\bea
g_{IJKL}[X^I,[X^J,X^K,X^L],Z]=0.
\eea
But because of the antisymmetry of $g_{IJKL}$, this is nothing but the (\ref{WFI}) form of the fundamental identity.

\subsection{Relation to Fuzzy 3-spheres}

What we have shown is that the BLG theory can in principle have BPS solutions corresponding to various calibrations. Now we show how the 3-algebra structure  incorporates the known fuzzy 3-sphere solutions. 
For a given calibration $g_{IJKL}$, we get a solution to (\ref{Nahm}) if we can solve the 3-algebra equation
\bea
\label{3BPS}
\frac{1}{6}g_{IJKL}[A^J,A^K,A^L]=A^I,
\eea
because then $X^I(\sigma)=f(\sigma)A^I$ is a solution to (\ref{Nahm}) for $f(\sigma)$ satisfying $\partial_\sigma f(\sigma) = f^3(\sigma)$.
If we imagine that the M5-branes are at $X^2(=\sigma)=0$, then $f(\sigma)=\frac{i}{\sqrt{2\sigma}}$. 

Now we show that the previous solutions \cite{Basu, bermcop} can all be seen as specific fuzzy 3-sphere realizations of the 3-algebra equation (\ref{3BPS}) for the case when the 3-algebra is ${\cal A}_4$. We first notice from earlier work that the fuzzy 3-sphere coordinates defined in terms of
matrices\footnote{See \cite{Ramgoolam} for the details of the construction and the definitions of the matrices $G^*, G^a$.} $G^a (a=1,2,3,4)$ furnish a representation of the 3-algebra ${\cal A}_4$ if one takes the 3-bracket to be defined by (upto some constant scalings which we ignore)
\bea
[G^a,G^b,G^c] \equiv [[G^*,G^a,G^b,G^c]] \sim \epsilon^{abcd} G^d  
\eea
Here the 4-bracket operation is the one used by \cite{Basu, bermcop} and is defined through ordinary matrix multiplication as the Nambu 4-bracket: $[[A_1,A_2,A_3,A_4]]=\sum_{{\rm permutations} \
\sigma}{\rm sgn}(\sigma)A_{\sigma_1}A_{\sigma_2}A_{\sigma_3}A_{\sigma_4}$.
Once we have such an explicit representation of ${\cal A}_4$, we can use it as a building block, and form various direct sums of these matrices to construct solutions of (\ref{3BPS}) and this is the standard recipe of \cite{bermcop}. Each block gives rise to a single copy of the fuzzy 3-sphere funnel. So the full configuration corresponds to many expanding fuzzy 3-spheres each giving rise to an M5-brane: the fact that the 3-spheres intersect results in the calibrated intersection of M5s.

Before we end this section, we mention that the relevance of the dimensionality 
of the fuzzy 3-sphere representation here has to be better understood. Naively, if one 
would translate the BIon intuition for fuzzy 2-spheres to the M2-M5 system, one 
would expect that the sum of the dimensions of the fuzzy 3-sphere representations 
in each block should give rise to the total number of M2 branes. But since BLG 
theory for the 3-algebra ${\cal A}_4$ has no free parameters (apart from the level of the Chern-Simons piece) analogous to the rank 
of the gauge group, this interpretation is far from obvious. In particular, the work of \cite{Jacques, Christopher} seems to suggest that the theory corresponds to two M2 branes. The fuzzy funnel picture and this picture seem to be at odds. Maybe the mystery can be resolved by a better understanding of the algebraic structures on M2-branes and their representation theory. A reference which tries to address the issue of number of degrees of freedom of fuzzy 3-spheres in this context is \cite{bermcop2}.

\section{Signature of the Trace Form}

The assumption that we made\footnote{If we believe the picture that M2-branes have to end on M5-brane configurations, the 3-algebra seems forced to be ${\cal A}_4$. One way to see this is to look at the target space of the boundary self-dual string of the BLG theory with boundary. One sees that it is 5+1 
dimensional only if the 3-algebra structure constants are those of ${\cal 
A}_4$ \cite{boundary}. But perhaps the generic non-positive 3-algebras correspond to something more exotic.} during the Bogomol'nyi 
construction, namely that the trace is positive definite in the 3-algebra 
is an incredibly strong one. In fact, it is shown in \cite{BPG} that the
only positive definite finite-dimensional 3-algebra with fully anti-symmetric structure 
constants of this form is ${\cal A}_4$ (and its disconnected copies). Fortunately, as 
we saw in the last section, this was enough to incorporate all the M2-brane funnel solutions in the literature. 

It turns out that the constraint relations (\ref{constraint}) can be obtained without relying on the Bogomol'nyi trick. The BPS equations 
arise from setting the SUSY variation $\delta \Psi=0$. Therefore an 
equation linear in derivatives which preserves some combination of the 
supersymmetries has to take the form (\ref{Nahm}). One can convince 
oneself that this is the case by looking at specific configurations of 
branes (like the ones considered in \cite{bermcop}) that leave specific 
supersymmetries unbroken and imposing the resulting relations on $\delta 
\Psi=0$. Once we accept that the BPS conditions are given by (\ref{Nahm}), 
one has to check that they are consistent with the equations of motion. 
From (\ref{EOMS}), we see that the relevant EOM takes the form 
$\partial^2 X^I_a-\frac{\partial V}{\partial X^{Ia}}=0.$ 
Taking a derivative of the BPS equation, getting rid of the derivatives, 
and finally taking a trace, one ends up precisely with the constraint 
relation (\ref{constraint}). In this derivation we never needed to use the positivity of the norm.

But of course, the constraints are not the whole story. Lets restrict our attention to 
{\em classical, static} solutions.
If we demand that a state is BPS, then it means that it satisfies $Q 
|\psi\rangle =0$ for some supercharge. This in turn means that $\langle 
\psi| \{Q, Q\} |\psi \rangle \sim \langle\psi| H|\psi \rangle =0$ in any 
positive definite Hilbert space. The energy of any state in a 
supersymmetric theory is non-negative, but it seems like the 
Hamiltonian written earlier (\ref{H}) is unbounded below if the trace is not positive definite. So the theory is pathological, at least classically.

Recently, many more solutions to the fundamental identity were constructed with a norm that was Lorentzian \cite{ABC}, by simply augmenting the structure constants of classical Lie algebras. Interestingly, these theories do not suffer from the problem we mentioned above. The scalar part of the Hamiltonian density for these theories can be written as 
\bea
\mathcal{H} = \frac{1}{2}{\rm Tr}\Big(\partial_{\sigma}X^I \partial_{\sigma}X^I\Big) - \partial_{\sigma}X^I_+\partial_{\sigma}X_-^I + \frac{1}{12} {\rm Tr}\Big(X_+^I [ X^J, X^K] + X^J_+ [ X^K, X^I ] + X_+^K [ X^I , X^J ]\Big)^2 
\eea
where the troublesome negative norm direction has been split off in a ``light-cone" notation. The interesting point is that after an integration by parts, $X^I_-$ shows up only as a Lagrange multiplier enforcing the constraint  $\partial_\sigma^2 X^I_+=0$. If we solve this with $X^I_+ \sim \sqrt{\lambda}$, where $\lambda$ is a positive parameter, then the Hamiltonian is schematically that of a $\lambda \phi^4$ theory. Since $\lambda \sim (X^I_+)^2$, we have a positive definite Hamiltonian. 
It would be interesting to explore these theories along the lines considered here.


\section* {\bf Acknowledgments}

It is a pleasure to thank David Berman for a set of lectures at KU Leuven 
that inspired us to think about M-theory branes and also for encouragement and advice through emails during the initial stages of this project. We also thank Giuseppe Milanesi for discussions. This work is supported in part by IISN - Belgium (convention 4.4505.86), by the Belgian National Lottery, 
by the European Commission FP6 RTN programme MRTN-CT-2004-005104 in which 
the authors are associated with V. U. Brussel, and by the 
Belgian Federal Science Policy Office through the Interuniversity 
Attraction Pole 
P5/27.

\section* {\bf Appendix}

Some ways to rewrite the fundamental identity are collected here,
\begin{eqnarray}
f^{[abc}{}_g f^{e]fg}{}_d &=& 0 \,, \label{WFI} \\
f^{abc}{}_g f^{efg}{}_d &=& 3f^{ef[a}{}_g f^{bc]g}{}_d.
\end{eqnarray}
The equivalence of (\ref{WFI}) to the fundamental identity is shown in Gran, Nilsson and Petersson \cite{Christopher}. Elementary, but useful symmetry relations are
\begin{eqnarray}
[mkl]&=&\frac{1}{3}(m[kl]+l[mk]+k[lm]) \\
\,[abcde]&=&\frac{1}{5}([abcd]e+[eabc]d+[deab]c+[cdea]b+[bcde]a),
\end{eqnarray}
where square brackets stand for anti-symmetrization.

\bibliographystyle{unsrt}

\end{document}